\documentclass[usenatbib]{mn2e}

\usepackage[dvips]{graphicx}
\usepackage{amssymb}
\usepackage{txfonts}

\newcommand{\integral}{{\textit{INTEGRAL}}}
\newcommand{\xte}{{\textit{RXTE}}}

\newcommand{\fermi}{{\textit{Fermi}}}
\newcommand{\agile}{{\textit{AGILE}}}
\newcommand{\swift}{{\textit{Swift}}}

\newcommand{\phisup}{\Phi}
\newcommand{\psup}{P_{\rm sup}}
\newcommand{\porb}{P_{\rm orb}}
\newcommand{\g}{$\gamma$}

\title[The doubling of the superorbital period of Cyg X-1]{The doubling of the superorbital period of Cyg X-1}

\author[A. A. Zdziarski, G. G. Pooley and G. K. Skinner]
{Andrzej A. Zdziarski,$^1$\thanks{E-mail: aaz@camk.edu.pl} Guy G. Pooley$^2$ and Gerald K. Skinner$^{3,4}$\\
Centrum Astronomiczne im.\ M. Kopernika, Bartycka 18, 00-716 Warszawa, Poland\\
$^2$Cavendish Laboratory, J. J. Thomson Avenue, Cambridge CB3 0HE\\
$^3$Astroparticle Physics Laboratory, Code 661, CRESST and NASA/Goddard Space Flight Center, Greenbelt, MD 20771, USA\\ 
$^4$Department of Astronomy, University of Maryland, College Park, MD 20742, USA\\
}

\date{Accepted 2010 November 15. Received 2010 November 15; in original form 2010 September 15}

\pagerange{\pageref{firstpage}--\pageref{lastpage}}
\pubyear{2010}

\begin{document}

\maketitle

\label{firstpage}

\begin{abstract}
We study properties of the superorbital modulation of the X-ray emission of Cyg X-1. We find that it has had a stable period of $\sim$300 d in soft and hard X-rays and in radio since 2005 until at least 2010, which is about double the previously seen period. This new period, seen in the hard spectral state only, is detected not only in the light curves but also in soft X-ray hardness ratios and in the amplitude of the orbital modulation. On the other hand, the spectral slope in hard X-rays, $\ga 20$ keV, averaged over superorbital bins is constant, and the soft and hard X-rays and the radio emission change in phase. This shows that the superorbital variability consists of changing the normalization of an intrinsic spectrum of a constant shape and of changes of the absorbing column density with the phase. The maximum column density is achieved at the superorbital minimum. The amplitude changes are likely to be caused by a changing viewing angle of an anisotropic emitter, most likely a precessing accretion disc. The constant shape of the intrinsic spectrum shows that this modulation is not caused by a changing accretion rate. The modulated absorbing column density shows the presence of a bulge around the disc centre, as proposed previously. We also find the change of the superorbital period from $\sim$150 d to $\sim$300 d to be associated with almost unchanged average X-ray fluxes, making the period change difficult to explain in the framework of disc-irradiation models. Finally, we find no correlation of the X-ray and radio properties with the reported detections in the GeV and TeV \g-ray range. 
\end{abstract}

\section{Introduction}
\label{intro}

A major characteristic time scale in X-ray binaries is their orbital period, which, in many cases, is seen in their light curves as periodic modulations, see, e.g., \citet{wen06} for a list of orbital periodicities from the All-Sky Monitor (ASM) aboard {\it Rossi X-ray Timing Explorer\/} (\xte; \citealt*{brs93,levine96}). The observed flux modulations appear due to a number of different physical effects, see \citet*{pzi08} (hereafter PZI08) for a list of the possibilities. 

In addition, a number of X-ray binaries show modulation at quasi-periods much longer than their orbital periods. Those long-period modulations, listed, e.g., in \citet{wen06} and \citet{ogdu01}, hereafter OD01, are called superorbital (hereafter SO). In most of the known cases, the SO periodicity (or quasi-periodicity) appears to be caused by precession of an accretion disc and/or jet, which either results in variable obscuration of emitted X-rays as in Her X-1 \citep{k73,roberts74}, or in changes the viewing angle of the presumed anisotropic emitter, as in SS 433 \citep{k80} or Cyg X-1 (e.g., \citealt*{l06}, hereafter L06, \citealt*{i07}, hereafter IZP07), or both. In some relatively rare cases, the SO periodicity is caused by modulation of the accretion rate, e.g., in the low-mass X-ray binaries 4U 1820--303 \citep*{z07a} and 4U 1636-536 \citep*{shih05,fbs09}.

\begin{figure*}
\centerline{\includegraphics[width=17.5cm]{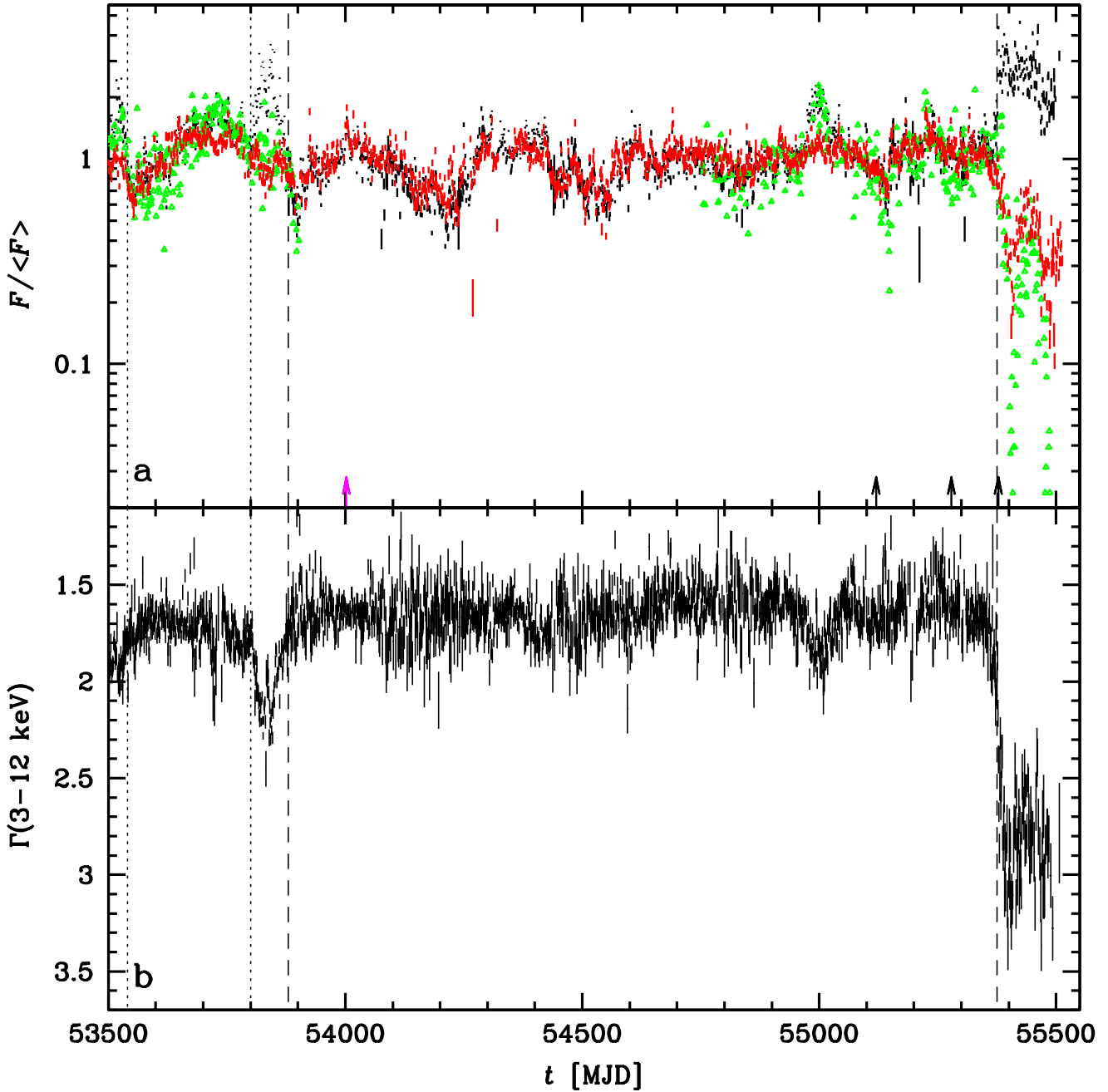}} 
\caption{
(a) The light curves starting from MJD 53500 showing the daily averages from ASM (black, 1.5--12 keV), BAT (red, 15--50 keV), and the Ryle/AMI radio telescope (green, 15 GHz), normalized to their respective average values within the hard state of MJD 53880--55375 (delineated by the vertical dashed lines). For clarity, the uncertainty of the radio points is not shown. The vertical dotted lines delineate the previous hard state, MJD 53540--53800. The arrows show the times of the reported \g-ray detections by MAGIC (magenta) and \agile\/ (black). (b) The corresponding 3--12 keV photon spectral indices, determined using the ASM data. The data at MJD $>55375$ show the 2010 soft state of Cyg X-1.
}
\label{f:lc}
\end{figure*}

In particular, Cyg X-1, a well-known high-mass X-ray binary containing a persistent accreting black-hole source (see, e.g., \citealt{zi05} for the properties of the binary), has shown such (quasi) periodicity in its X-ray, optical and radio emission with an average SO period of $\psup\simeq 150$ d, found in a large number of investigations (\citealt*{brocksopp99a,poo99,kitamoto00,kar01,od01,benlloch01,benlloch04}; L06; IZP07). The modulation is detectable in the hard X-ray spectral state (see, e.g., \citealt{zg04}), in which Cyg X-1 spends most of the time. No SO modulation has been detected in the soft states, but this seems to be due to their typical short duration and strong overall variability. The underlying physical process appears to still operate, as indicated by the reappearance of the SO modulation at a phase expected from past modulation at a next hard state (L06; IZP07).

The reality of the $\sim$150-d period was then questioned by \citet{rico08}, hereafter R08. Although he did find the strongest, and highly significant, power-spectrum peak at $148\pm 2$ d using data similar to those analysed by IZP07 (who found $\psup\simeq 152$ d), he claimed it to be ``an artefact of Fourier-power-based analysis". The basis for this claim was his finding of a variable $\psup$ in three separate sub-samples each covering $\simeq$300 d. We disagree with this rejection of the reality of the 150-d peak. First, determination of the modulation period based on a single interval covering $\simeq 2\psup$ is obviously questionable. Although R08 claimed those individual periods were highly significant statistically, he determined their probability under the assumption of the presence of white noise only, which is well known to strongly overestimate the actual significance in astrophysical sources (e.g., \citealt{is96}). Second, unlike the periodic orbital modulation, the SO modulation represents a quasi-periodic modulation, in which the underlying clock is not perfect, and where the period changes from one cycle to another. Furthermore, the presence of an aperiodic long-term variability, related to the variable stellar wind from the companion (which also causes spectral state changes in Cyg X-1, e.g., \citealt{gies03}), and possible accretion disc instabilities, obviously distorts the SO period (presumably related to precession) if measured in a short interval. 

Thus, the 150-d period does represent the {\it average\/} $\psup$ as measured from $\sim 1980$ to 2004 (L06; IZP07). The physical reality of this average period was then strongly confirmed by PZI08, who found that both the X-ray spectral hardness and the depth of the X-ray orbital modulation are modulated with this $\psup$.

On the other hand, R08 found a striking change of the SO period to $\psup\ga 300$ d occurring in 2005. The data at his disposal, until 2008, covered only $\sim 3\psup$, during which the period was $\simeq 320$--330 d. Notably, the new period is similar to that, $294\pm 4$, seen in the optical and X-ray ranges from 1969 to $\sim$1980 \citep*{pth83,kemp83}, and apparently in the optical data till 1985 \citep{kemp87}. 

In the present work, we find the $\sim$300-d period to persist until the recent switch to the soft state in 2010 June \citep{negoro10,rushton10}. We find the new $\psup\sim$300-d modulation to have the same physical properties as those found by PZI08. Namely, it appears not only in the soft and hard X-ray and radio light curves, but also in the phase dependence of the X-ray hardness and of the depth of the X-ray orbital modulation. Furthermore, we find very strong evidence that the SO modulation is due to changes of the source geometry but not due a variable accretion rate. 

\section{The data}
\label{data}

We study monitoring data from the \xte/ASM, which contain energy channels at 1.5--3 keV, 3--5 keV and 5--12 keV. We apply a small correction for the empty-field ASM count rates of 1 mCrab, appropriate for Cyg X-1 (R. Remillard, personal communication). We then use the monitoring by the Burst Alert Telescope (BAT, \citealt{barthelmy05,m05}) on board the \swift\/ satellite. We use both the 15--50 keV data from the BAT web page\footnote{http://swift.gsfc.nasa.gov/docs/swift/results/transients} and 14--195 keV 8-channel light curves created for this work. The channels are between energies of 14, 20, 24, 35, 50, 75, 100, 150 and 195 keV. 

The ASM count rates are converted into energy fluxes using the matrix of \citet{z02}, hereafter Z02. We have checked that the fluxes obtained are within 10 per cent of those obtained by scaling the count rates to those from the Crab and using the best fit to the 0.5--180 keV Crab energy spectrum to data from several detectors of \citet{rv01}. We convert the BAT 8-channel data into energy fluxes by scaling to the Crab spectrum. We use the results of \citet{rv01} up to 100 keV, and at higher energies we use those of \citet{jr09}, based on the \integral/SPI data. This gives us,
\begin{equation}
F(E)=A_i \exp(-N_{\rm H}\sigma) (E/1\,{\rm keV})^{-\Gamma_i+1},
\label{crab}
\end{equation}
where the H column density is $N_{\rm H}=3.23\pm 0.02$ cm$^{-2}$, $\sigma$ is the bound-free cross section per H atom averaged over the cosmic composition, $A_1=10\pm 1$ cm$^{-2}$ s$^{-1}$, $\Gamma_1=2.10\pm 0.01$ for $E\le E_{\rm b}=100$ keV, $\Gamma_2=2.24\pm 0.02$ at $E>E_{\rm b}$, and $A_2$ follows from the continuity at $E_{\rm b}$. 

In addition, we use 15-GHz data from the Ryle Telescope and the AMI Large Array. The AMI Large Array is the re-built and reconfigured Ryle Telescope. The monitoring by Ryle was carried out until 2006 June 16, and the AMI monitoring of Cyg X-1 started 2008 April 17. The new correlator has a bandwidth of about 4 GHz (compared with 350 MHz); but the effective centre frequency is similar, and in any case the radio spectrum of Cyg X-1, at least in the hard state, is known to be very flat \citep{fender00}. The data are subject of variations in the flux calibration of about 10 per cent from one day to another.

We take into account the barycentric correction for all the data. We first investigate data from the time interval considered by R08 but extended until 2010. Also, we study the uninterrupted long hard state of 2006--2010, which lasted $\sim$1500 d.

\begin{figure}
\centerline{\includegraphics[width=\columnwidth]{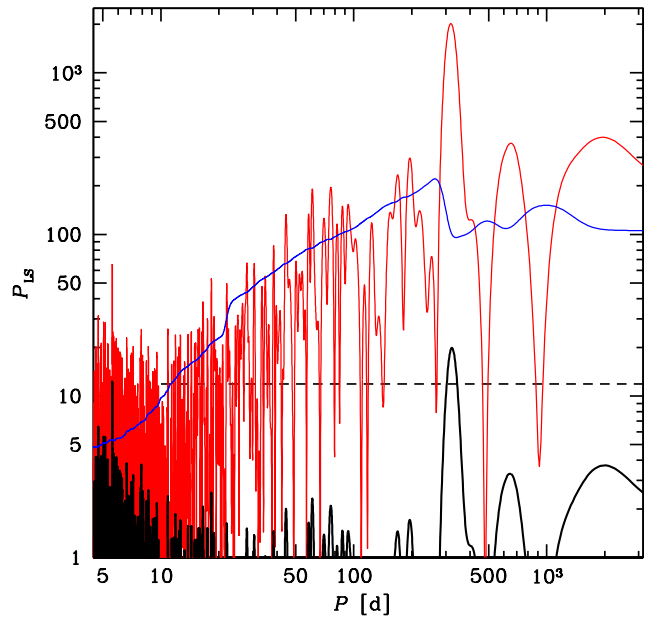}} 
\caption{The Lomb-Scargle periodograms for the ASM 5--12 keV dwell data (red curve) obtained using the hard-state intervals of MJD 53540--53800 and 53880--55375. The horizontal dashed line shows the level of the 90 per cent significance for $p_N$. We see many apparently highly significant peaks. The black curve shows the ASM periodogram divided by the one describing the red noise (blue curve) using the method of \citet{is96}, see Section \ref{lc}. Now only the $\sim$300 d and the (orbital) 5.6-d peaks are significant.
}
\label{f:period}
\end{figure}

\begin{figure*}
\centerline{\includegraphics[width=\columnwidth]{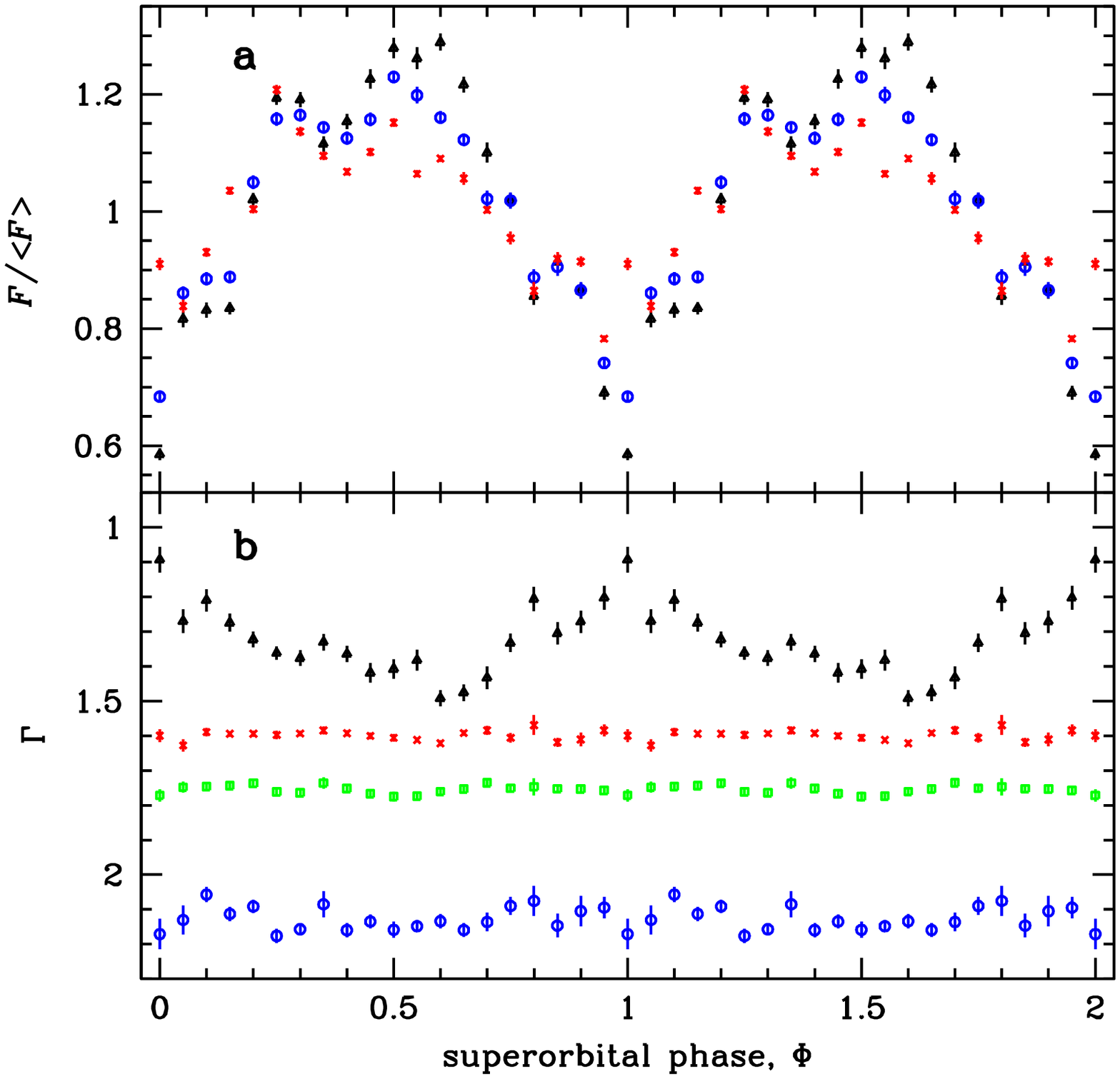}\hbox to 0.3cm{\hfill}
\includegraphics[width=\columnwidth]{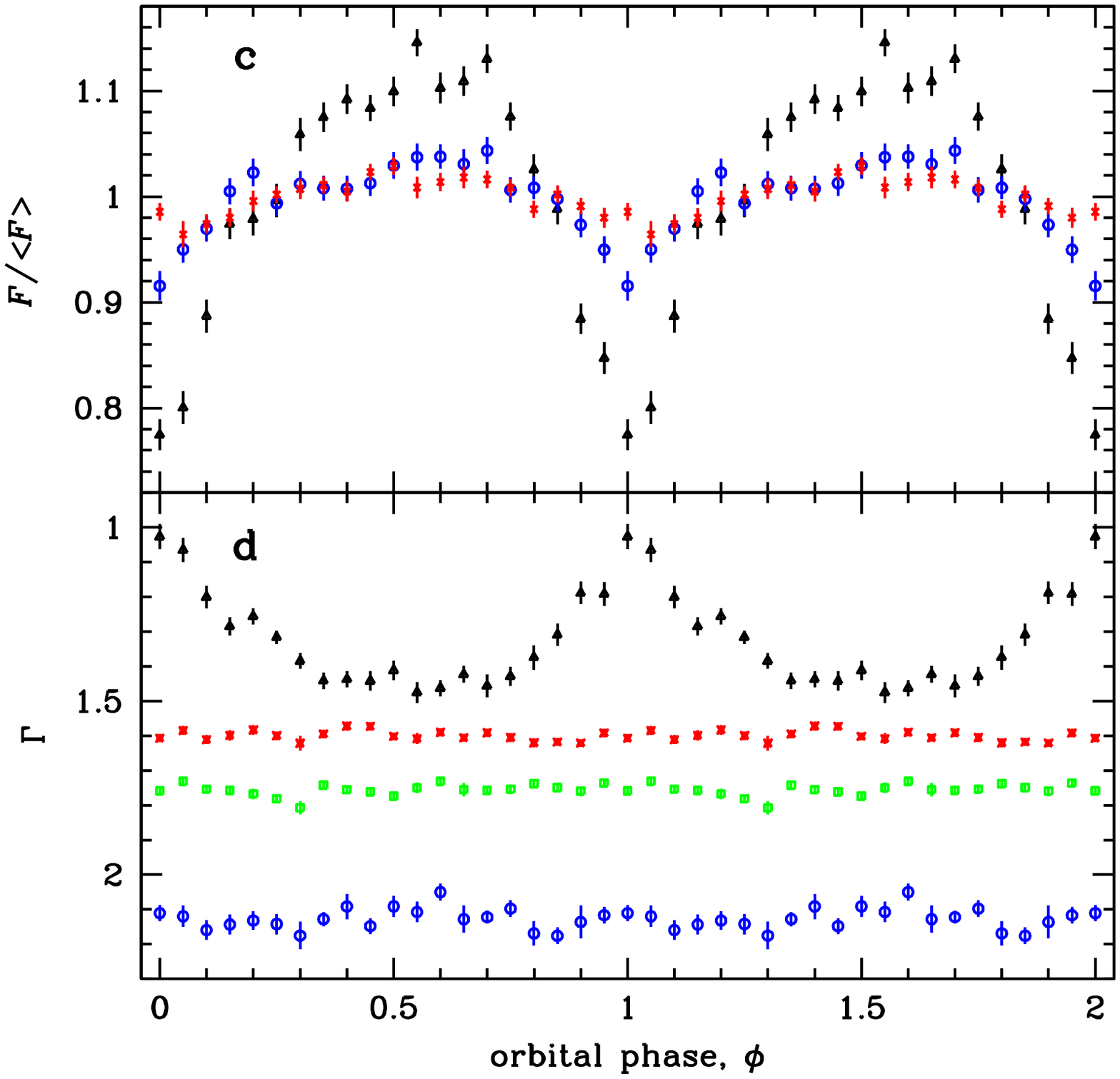}} 
\caption{(a) The MJD 53880--55375 light curves from ASM at 1.5--3 keV (black triangles), 5--12 keV (blue circles) and BAT at 15--50 keV (red crosses) folded over the SO period for the ephemeris of equation (\ref{ephemeris}). The dependencies are normalized to their respective average values. (b) The corresponding photon spectral indices in the energy ranges of 1.5--5 keV (black triangles), 20--35 keV (red crosses), 35--75 keV (magenta squares) and 75--150 keV (blue circles). (c, d) The light curves and spectral indices as in (a, b), respectively, but folded over the orbital period. 
}
\label{f:folded}
\end{figure*}

\section{Results}
\label{results}

\subsection{The light curves and power spectra}
\label{lc}

Fig.\ \ref{f:lc}(a) presents the daily-average light curves from ASM (1.5--12 keV), the BAT transient monitoring (15--50 keV), and the Ryle/AMI 15-GHz monitoring. They have been renormalized to their respective average values during the long 2006--2010 hard state, MJD 53880--55375. The (un-weighted) averages are 20.6 s$^{-1}$, 0.173 s$^{-1}$, and 12.7 mJy, respectively. The corresponding rms, i.e., the fractional intrinsic standard deviation (after subtraction of the contribution from the measurement errors), is 0.27, 0.19 and 0.35, respectively. 

We see that the fluxes from the two X-ray instruments closely follow each other during the 2006--2010 hard state as well as the previous one, MJD 53540--53800. The radio flux, when available, strongly correlates with the ASM flux in the hard state. (We discuss the radio/X-ray correlation in Cyg X-1 in detail in \citealt{z11}.) We also calculate an equivalent spectral index for each pair of adjacent energy channels using the method of Z02, see Appendix \ref{index} for details. This spectral index corresponds to a power law spectrum that would have the same energy flux in each of the channels as that of the data. During both hard states, the 3--12 keV photon spectral index was $\Gamma\sim 1.7$, and it was always (except for three isolated occurrences) $<2.1$, as shown in Fig.\ \ref{f:lc}(b). We see a significant increase of the ASM flux around MJD $\sim$55000, during which, however, the source remained in the hard state, as shown by $\Gamma\la 2$ and the correlation of the ASM and the radio fluxes. 

Fig.\ \ref{f:lc} also shows (arrows) the times when the Major Atmospheric Gamma Imaging Cherenkov Telescope (MAGIC) \citep{albert07} and \agile\/ \citep{sabatini10a,sabatini10b,bulgarelli10} reported detections of Cyg X-1 in TeV and GeV respectively. We see that these events do not seem to correlate with any source properties in X-rays, except for the last detection by \agile, which took place around the transition to the soft state in 2010 June. All of the \agile\/ detections occurred in rather low radio states.

In Fig.\ \ref{f:lc}, we also clearly see the $\sim$300-d modulation (excluding the soft states), covering now 6 periods. The significance of this periodicity can be quantified using the periodogram of \citet{lomb} and \citet{scargle}. The periodogram obtained using the ASM 5--12 keV (which channel is least affected by the orbital modulation) dwell data is shown by the red curve in Fig.\ \ref{f:period}. The periodogram for the BAT 15--50 keV orbital data is similar, and, for brevity, it is not shown. Under the assumption of the presence of periodicities and white noise only, the probability of finding a periodicity by chance in one of $N$ trials is, 
\begin{equation}
p_N =1-(1-p_1)^N,
\label{pn}
\end{equation}
where $p_1=\exp(-P_{\rm LS})$ is the probability for a single trial, $P_{\rm LS}$ is the Lomb-Scargle power and the number of independent frequencies, $N$, approximately equals 1/2 of the number of the independent data points. Using this definition, we would have obtained many periodicities with {\it extreme\/} significance. This was, in fact, done by R08, who claimed, e.g., the chance probability of $10^{-96}$ for the $\sim$300-d peak in his ASM periodogram (obtained using daily averages, and until MJD 54592).

However, it is well-known that power spectra of X-ray binaries, and in particular, Cyg X-1, show quite strong components at low frequencies, and are thus not at all compatible with white noise, see, e.g., \citet{rpk02}. This effect, often called red noise, can be taken into account, e.g., by averaging the periodogram over frequencies surrounding a given peak, and then dividing the original Lomb-Scargle power by the average one \citep{is96}. We use here this method as implemented in \citet{wen06}, see that paper for details. The corrected probabilities are now given by the same formula as above, but using the rescaled $P_{\rm LS}$, which is shown by the black curve in Fig.\ \ref{f:period}. For the ASM and BAT data, the rescaled peak power has the values of $P_{\rm LS}\simeq 20$, 16.5, and is at $\psup\simeq 319\pm 5$ d, $312\pm 5$ d, respectively, where the uncertainty has been calculated using eq.\ (3) of \citet{wen06}. For the respective $N= 15174$ and 10462, we find $p_N\simeq 3\times 10^{-5}$, $7\times 10^{-4}$. We note that these are conservative estimates since the number of independent data points is significantly lower (and the corresponding probability lower) because the aperiodic components of light curves clearly exhibit a degree of autocorrelation, and thus adjacent measurements within the autocorrelation length are not independent, see, e.g., L06. However, since we have already established the significance of the periodicity at a very high level, we ignore this complication. Note that we do not find any peak at $\psup\simeq 1000$ d, which was found by R08 and speculated to be real, but which appears to simply reflect the total duration of his data set.

Given the 2006--2008 gap in the radio coverage, the power spectrum of the radio data has a rather broad SO peak, and we do not study it in detail. Still, the maximum LS power corresponds to $\psup=316$, in agreement with those of the ASM and BAT data. 

The full data set considered above suffers from being interrupted by the occurrence of the soft state during MJD 53800--53880, which appears to have introduce some phase shift into the SO modulation resulting in an apparent lengthening of the first cycle, see Fig.\ \ref{f:lc}. On the other hand, the data sets for MJD 53880--55375 offer us an opportunity to study a very long, uninterrupted, hard state covered by both the ASM and BAT as well as partly by the radio monitoring. The periodograms for MJD 53880--55375 look similar to those shown in Fig.\ \ref{f:period}, and thus we do not show them here. We find $\psup=315\pm 6$ d, $306\pm 6$ d for the ASM 5--12 keV and BAT 15--50 keV data, respectively. 

\subsection{Folded light curves}
\label{folded}

For the MJD 53880--55375 data set, we now search for the dependence of the X-ray hardness and the depth of the orbital X-ray modulation on the SO phase, which effects were earlier found by PZI08 for the $\sim$150-d SO modulation. We average now the logarithms of the count rates (without weights), which is motivated by their normal distributions, whereas the count rates themselves have the distributions far from normal (PZI08). The uncertainties on the averages are calculated from the dispersion of the individual measurements (without taking into account their errors). As above, we use the ASM dwell and BAT orbital data. The zero SO phase, $\phisup=0$ is taken to correspond to the minimum ASM count rate. To facilitate the study of the dependence of the orbital modulation on the SO phase, we take a value of the SO period within the uncertainties of the periods found in the data but equal to a multiple of the orbital period, $\porb=5.6$ d. This results in the ephemeris,
\begin{equation}
{\rm min[MJD]} = 53275.8+313.6 E,
\label{ephemeris}
\end{equation}
where $E$ is an integer. The resulting folded curves of the count rates are shown in Fig.\ \ref{f:folded}(a). We see that the amplitude of the SO modulation decreases with energy. The amplitude at 1.5--3 keV is $0.6 \la F/\langle F\rangle \la 1.3$ whereas it is only about $\pm 0.1$ in the 15--50 keV range. This is compatible with the modulation being enhanced by bound-free absorption around the flux minima (as found by PZI08 for the $\sim$150-d modulation), as well as with the degree of anisotropy of Compton-scattered spectra decreasing with photon energy, see IZP07. Fig.\ \ref{f:folded}(b) shows the corresponding spectral hardness as given by the photon spectral index calculated using the method of Z02. We see that indeed the 1.5--5 keV spectra become significantly harder at the minimum of the SO modulation. On the other hand, we see no spectral modulation above 20 keV, as shown by three spectral indices in the 20--150 keV range. The SO modulation thus consists of a modulation of the total flux and of the column density only, but not of the slope of the intrinsic spectrum. This is fully compatible with the modulation being caused by changing the viewing angle of the precessing accretion disc, but not by changing the accretion rate. 

Compared to the previous hard states covered by the ASM, from MJD 50350 to 53174 (IZP07; PZI08, during which period was $\psup\simeq 150$ d), the present SO modulation is significantly stronger, especially the minima caused by absorption are deeper. The SO modulation of $\Gamma(1.5$--5) keV is also much stronger now, as compared to its variability within $\Gamma\simeq (1.3$--1.5) during the previous hard states.

On the other hand, we find the ASM count rates averaged over the hard states characterized by $\psup\simeq 150$ d and $\psup\simeq 300$ d to be to be almost identical. The former in the 1.5--3 keV, 3--5 keV and 5--12 keV bands are higher than the latter by only 7, 2 and 3 per cent, respectively. The largest difference being in the 1.5--3 keV rate is compatible with the found increased absorption during the $\psup\simeq 300$-d epoch, whereas the rates at to higher energies (corresponding to the intrinsic emission and not significantly affected by absorption) are practically identical. Furthermore, the bolometric fluxes calculated using the ASM and {\it Compton Gamma Ray Observatory\/} BATSE data, and the ASM and BAT data, which correspond to these two respective epochs, are also almost identical, as found by \citet{z11}.

Figs.\ \ref{f:folded}(c,d) show the same light curves and indices as in (a,b) but folded over the 5.6 d orbital period, for which the ephemeris of \citet{lasala98} and \citet{brocksopp99b} has been used (see eq.\ [1] in IZP07). Overall, the orbital modulations are quite similar to the SO ones. The main difference is the relative weakness of the 15--50 keV modulation (though still significant), now about $\pm 2$ per cent. This modulation is also found in the periodogram, which has the rescaled $P_{\rm LS}\simeq 6.5$ at $P_{\rm orb}=5.6$ d, for which the single-trial probability (appropriate for modulation at a {\it known\/} frequency, see, e.g., L06) is $p_1\simeq 0.0015$. These results are similar (though more significant) to those found by L06 for the BATSE 20--100 keV data. The modulation is explained (L06) as being due to Compton scattering of hard X-rays away from the line of sight by the same gas (of the stellar wind of the donor star) as that causing the orbital modulation of soft X-rays due to bound-free absorption \citep{wen99}. Again, we see no modulation of the spectral indices at $E\ge 20$ keV, which is explained by the weakness of the bound-free absorption and constancy of the Compton cross section at those energies. Compared to the previous hard states (MJD 50350--53174), the present orbital modulation is significantly stronger, e.g., the minimum of the 1.5--3 keV relative flux was before about 0.85, and $\Gamma(1.5$--5 keV) was within 1.3--1.6 only.

\begin{figure}
\centerline{\includegraphics[width=\columnwidth]{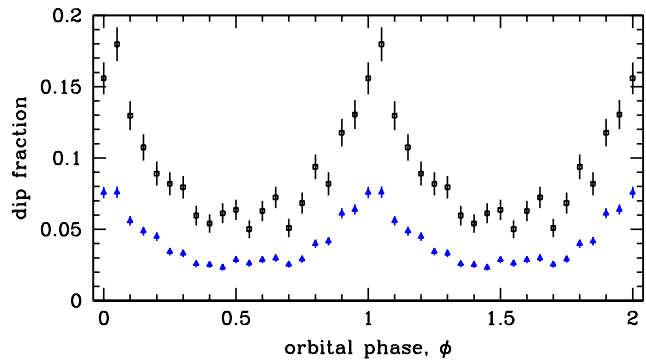}} 
\caption{The distribution of the fractional frequency of occurrences of X-ray dips (see Section \ref{folded}) over the orbital phase during MJD 53880--55375 (black squares) and the entire currently available ASM data, MJD 50087--55406 (blue triangles). The uncertainties shown assume Poisson's statistics. 
}
\label{f:dip}
\end{figure}

\begin{figure}
\centerline{\includegraphics[width=\columnwidth]{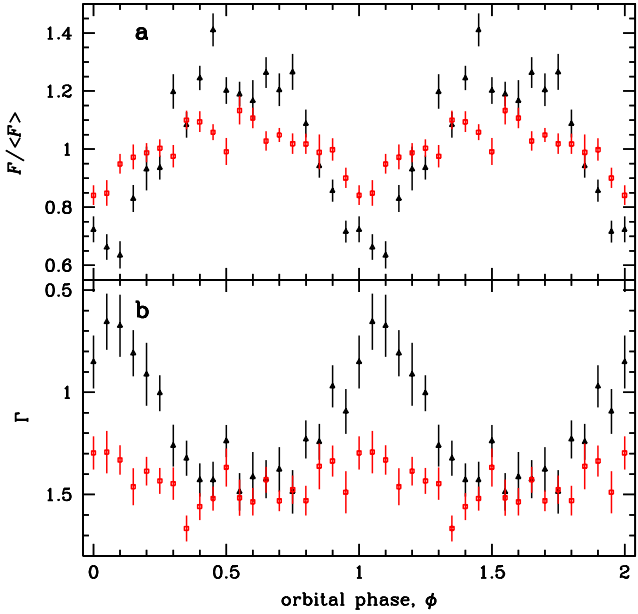}} 
\caption{(a) The MJD 53880--55375 light curves from the ASM 1.5--3 keV folded over the orbital period and for a chosen interval of the SO phase of equation (\ref{ephemeris}): $\Phi=-0.05$ to +0.05 (black triangles), and 0.55--0.65 (red squares). The dependencies are normalized to their respective average values. (b) The corresponding 1.5--5 keV spectral index. 
}
\label{f:orb_so}
\end{figure}

Another measure of the orbital X-ray modulation is the frequency of X-ray dips, defined by increases of the X-ray hardness by \citet{bal00}. They are presumed to originate during passage of clouds of the clumpy wind through the line of sight. \citet{bal00} defined them as events in the ASM dwell light curve with the ratio of counts of either (3--5 keV)/(1.5--3 keV) $>2$ or (5--12 keV)/(3--5 keV) $>2.5$. Fig.\ \ref{f:dip} shows the fractional occurrence of the dips as a function of the orbital phase. We see the dips are concentrated around the orbital phase 0, as found before (\citealt{bal00}; PZI08), with some shift of the peak towards $\phi>0$. We also found an SO dip modulation, with similar shape as in PZI08. Compared to the previous hard states, within MJD 50087--53789, studied by PZI08, we find the dips are now much more frequent. We find 2251 of them among 25147 of the data points, compared to 1151/31211 in the study of PZI08. In Fig.\ \ref{f:dip}, we also show the dip distribution for the entire currently available ASM data, with the frequency of 3929/96018 (similar to that of PZI08). The present increase of the dip frequency is compatible with the orbital modulation being stronger during the presently considered interval than at previous occurrences of the hard state (see above). In either case, we see no evidence of the secondary peak at $\phi\simeq 0.6$ claimed (based on data with rather low statistics) by \citet{bal00}, in agreement with the finding of PZI08. 

\begin{figure}
\centerline{\includegraphics[width=\columnwidth]{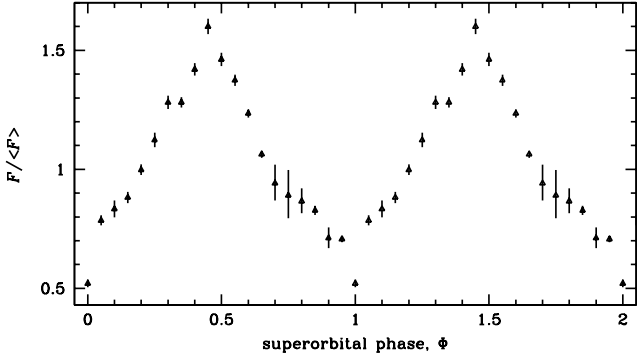}} 
\caption{(a) The MJD 53540--53800 and 53880--55375 light curves from the Ryle/AMI radio telescope folded over the SO period of $\psup= 313.6$ d.
}
\label{f:radio}
\end{figure}

PZI08 found that the orbital modulation is significantly stronger near the minimum of the SO cycle than near its maximum, and fitted its dependence using a model of a bulge around the centre of the precessing disc. Here, we confirm it for the present data (with the higher $\psup$). Fig.\ \ref{f:orb_so}(a) shows the profile of the orbital modulation for two ranges of the SO phase; around the SO minimum, $\Phi=(-0.05$ to +0.05), and near the maximum (see Fig.\ \ref{f:folded}a), $\Phi=(0.55$--0.65). We see that the latter profile is much shallower, $0.85 \la F/\langle F\rangle \la 1.1$, than the former, $\pm 0.4$. This is also very clearly seen in the modulation of the 1.5--5 keV spectral index, see Fig.\ \ref{f:orb_so}(b), again much stronger at the former range of $\Phi$ than the latter. 

Fig.\ \ref{f:radio} shows the SO modulation of the radio flux. Given the 2006--2008 gap in the radio coverage, we use now both of the occurrences of the hard state. We see a very significant modulation, again stronger than during the previous hard states (which relative amplitude was about 0.7--1.3). As discussed in IZP07, the most likely cause of the modulation is the precession of the jet tied to that of the disc. Then, the jet emits anisotropically due to the relativistic beaming. 

\section{Discussion and conclusions}
\label{discussion}

We have found that the SO period of $\sim 300$ d, which onset in 2005 was found by R08, has persisted in the X-ray and radio light curves of Cyg X-1 at least until the switch to the soft state of Cyg X-1 in 2010 June. The statistical significance of this periodicity, at $\psup\simeq 310$--320 d, corresponds to a chance probability of $<3\times 10^{-5}$. 

In addition to the X-ray and radio light curves, we find the SO modulation to also be very significantly present in the spectral hardness, or, equivalently, a local spectral index, in soft X-rays, $E\la 10$ keV. Also, the depth of the soft X-ray orbital modulation (caused by absorption in the stellar wind) is found to be a strong function of the SO phase, with the orbital modulation being much stronger around the minimum flux of the SO cycle than at its minimum. Other known cases of correlations between the orbital modulation and superorbital phase are those in the low-mass X-ray binary 4U 1820--303 \citep{z07b} and in some X-ray pulsars \citep{rp08}. In the latter case, this phenomenon appears also be due to a changing viewing angle of a precessing system.

On the other hand, thanks to the \swift/BAT data, we find a new feature of the SO modulation of Cyg X-1. Namely, the average spectral slope at $E\ga 20$ keV is completely constant with the SO phase. As discussed, e.g., in Z02, the long-term variability of Cyg X-1 in the hard state is driven mostly by variable irradiation of the hot, Compton-scattering plasma, by some soft photons, causing changes of the spectral slope and related to changing accretion rate. This type of variability is absent when averaged over the SO period. This provides strong evidence that the SO modulation is not driven by a variable accretion rate. The SO modulation consists of changes of the amplitude of the intrinsic spectrum, and of enhanced soft X-ray absorption around the minimum flux of the SO cycle. These features are fully compatible with the model of a bulge around the centre of a precessing disc, proposed by PZI08.

We note that the precession mechanism should be compatible with the change of its period by a factor of 2. This appears easier to achieve with radiation driven precession than with that caused purely by the tidal forces exerted by the donor (e.g., \citealt{wp99}; OD01; \citealt{caproni06}). OD01 find that Cyg X-1 lies in a region of marginal stability of radiation induced precession, which possibly can explain the observed large change of $\psup$ as a manifestation of a chaotic behaviour. Still, if this is the case, the rather coherent $\sim$150-d and $\sim$300-d SO modulations over time scales of years are difficult to understand. 

An alternative is that the SO modulation is not chaotic but stable with the period of $\sim$300 d, but there are changes in the nodal precession mode. The observed period of $\simeq 300$ d would then correspond to the main, retrograde mode with $m=0$ bending nodes (OD01). The $\sim$150-d observed period would then correspond to the prograde $m=1$ mode (OD01), in which the actual precession period would be still $\sim$300 d, but the disc bending gives rise to two maxima/minima of the observed flux per period. An argument in favour of this interpretation is that the precession during the $\sim$150-d modulation was found prograde by PZI08. In this model, the absorption periodically varying with a half of the precession period needs to be due to some medium around the bent disc.

We have found that the ASM count rates averaged over the hard states characterized by $\psup\simeq 150$ d and $\psup\simeq 300$ d are almost identical. The bolometric fluxes in these two epoch are also found to be almost identical \citep{z11}. Thus, the change of the SO period is not associated with a significant change of the average luminosity (most likely tied to the accretion rate), which appears to present a difficulty for the model of radiation-driven precession.

We also study the orbital modulation of the X-ray light curves. We find that the depth of this modulation in the soft X-rays is significantly higher during 2006--2010 than before. This also manifests itself in the frequency of the X-ray dips, much higher in this epoch than before. This appears to indicate that the stellar wind, for some reasons, has become more anisotropic than before. Likely, the bulge at the disc centre may have now a higher optical depth than before. 

We find statistically significant orbital modulation of the flux (but not of the spectral slope) in the \swift\/ BAT light curve. The cause of the modulation appears to be Compton scattering of photons away from the line of sight (L06). As the optical depth of the wind is higher at the superior conjunction than at the inferior one, this scattering causes changes of the observed hard X-ray flux by about $\pm 2$ per cent. 

Finally, we find no correlation of the X-ray and radio properties of the source with the reported detections in high energy \g-rays. The TeV detection happened during a relatively high X-ray flux (see also \citealt{malzac08}), whereas the GeV ones took place when both the X-ray and radio fluxes were low. We note that \fermi\/ LAT detected no GeV emission from Cyg X-1 \citep{hill10} during the first two flares reported by \agile\/ \citep{sabatini10a,bulgarelli10}. 

\section*{ACKNOWLEDGMENTS}

We thank A. Ogorza{\l}ek and L. Wen for help with calculating the rescaled periodograms, H. Krimm and P. Lubi{\'n}ski for help with the BAT data, G. Dubus for discussion of radiation induced precession, and the referee for valuable suggestions. This research has been supported in part by the Polish MNiSW grants NN203065933 and 362/1/N-INTEGRAL/2008/09/0. The AMI Arrays are operated by the University of Cambridge and supported by the STFC. We acknowledge the use of data obtained through the HEASARC online service provided by NASA/GSFC.

\appendix
\section{Calculation of the equivalent spectral indices}
\label{index}

Often, hardness of spectra is described by the ratio of count rates or fluxes, $R$, in two channels. However, even if energy or photon fluxes are used, this value depends on the width of the channels, and is just specific to a given instrument. On the other hand, a hardness ratio can be uniquely converted to the equivalent spectral index, $\Gamma$, corresponding to a power law spectrum with the same ratio of fluxes as the actual data. Thus, this gives a local spectral slope, which can usually be interpreted physically, unlike the hardness ratio, which is an abstract and instrument-dependent quantity. Note that this index is a unique and monotonous function of the flux hardness ratio. This index is analogous to colours used in stellar astrophysics, which also describe a spectrum in an instrument-independent way. This technique was introduced by Z02, who, however, gave no details of the calculation. We present them here.

The condition of the ratio of the energy fluxes in a power law
spectrum being equal to that in the observed spectrum, $R$, can be written as,
\begin{equation} 
R = \cases{\displaystyle {E_4^{2-\Gamma} -
E_3^{2-\Gamma}\over E_2^{2-\Gamma} - E_1^{2-\Gamma}}, & $\Gamma\neq 2$;\cr
\displaystyle {\ln{E_4/E_3}\over \ln{E_2/E_1}}, & $\Gamma= 2$,\cr} \label{eq:hr}
\end{equation}
where the channel boundary energies are $(E_1,\,E_2)$, $(E_3,\,E_4)$. For adjacent channels with increasing energy, $E_2=E_3$ (assumed in this work), which, however, is not required in this formalism. We define,
\begin{equation}
a\equiv {E_1\over E_2}<1,\quad b\equiv {E_4\over E_3}>1, \quad g\equiv 2-\Gamma.
\end{equation}
In order to find $\Gamma(R)$, we need to solve,
\begin{eqnarray}
\lefteqn{
f(g)=b^g+R a^g-1 -R=0,\quad g\neq 0;\label{f_g}}\\
\lefteqn{R=-{\ln b\over \ln a},\quad g=0.}
\end{eqnarray}
Equation (\ref{f_g}) can be readily solved by Newton's method, in which case the derivative is also needed,
\begin{equation}
f'(g) =a^g  R \ln a +b^g \ln b.
\end{equation}

The uncertainty of $\Gamma$ is given by $\Delta \Gamma = \Delta R / \vert {\rm d} R/{\rm d} \Gamma\vert$, where $\Delta R$ is the uncertainty of $R$. We find,
\begin{equation} 
\left\vert{{\rm d} R\over {\rm d} \Gamma}\right\vert = \cases{ \displaystyle
{a^g(b^g-1)\ln a + b^g (1-a^g) \ln b\over (1-a^g)^2}, & $\Gamma\neq 2$;\cr
\displaystyle{(\ln a-\ln b)\ln b\over 2\ln a}, & $\Gamma= 2$.\cr} \label{eq:drdp}
\end{equation}

If $R$ is given by the ratio of photon, rather than energy, fluxes, then $2-\Gamma$ is replaced by $1-\Gamma$ in equation (\ref{eq:hr}) and $g=1-\Gamma$. The special cases of $\Gamma=2$ are now for $\Gamma=1$.

\label{lastpage}

\end{document}